\documentclass[a4paper,11pt]{article}
\usepackage{jheppub} % for details on the use of the package, please
\usepackage[T1]{fontenc} % if needed

\newcommand{\ra}{\rightarrow}

\newcommand{\prl}{Phys.~Rev.~Lett.}
\newcommand{\prd}{Phys.~Rev.~{\bf D}}
\newcommand{\plb}{Phys.~Lett.~{\bf B}}
\newcommand{\etal}{{\textit{et al.}}}

\title{\boldmath  Search for solar axions with CsI(Tl) crystal detectors}

\author[a]{Y.S. Yoon,}
\author[a,b,1]{H.K. Park, \note{Corresponding author.}}
\author[c]{H.~Bhang,}
\author[c]{J.H.~Choi,}
\author[c]{S.~Choi,}
\author[e]{I.S.~Hahn,}
\author[a]{E.J.~Jeon,}
\author[c]{H.W.~Joo,}
\author[a]{W.G.~Kang,}
\author[c]{B.H.~Kim,}
\author[c]{G.B.~Kim,}
\author[f]{H.J.~Kim,}
\author[c]{K.W.~Kim,}
\author[c]{S.C.~Kim,}
\author[c]{S.K.~Kim,}
\author[a,b,d]{Y.D.~Kim,}
\author[a,b,g]{Y.H.~Kim,}
\author[a,b]{H.S.~Lee,}
\author[c]{J.H.~Lee,}
\author[c]{J.K.~Lee,}
\author[a]{D.S.~Leonard,}
\author[a]{J.~Li,}
\author[c]{S.S.~Myung,}
\author[a]{S.L.~Olsen,}
\author[a]{J.H.~So,}
\affiliation[a]{Center for Underground Physics, Institute for Basic Science (IBS), Daejon 34047, Korea}
\affiliation[b]{Basic Science, IBS-UST School, Daejeon 34047, Korea}
\affiliation[c]{Department of Physics and Astronomy, Seoul National University, Seoul 08826, Korea}
\affiliation[d]{Department of Physics, Sejong University, Seoul 05006, Korea}
\affiliation[e]{Department of Science Education, Ewha Womans University, Seoul 03760, Korea}
\affiliation[f]{Department of Physics, Kyungpook National University, Daegu 41566, Korea}
\affiliation[g]{Korea Research Institute of Standards and Science, Daejon 34113, Korea}
\affiliation[h]{Department of Engineering Physics, Tsinghua University, Beijing 100084, China}
\affiliation[i]{Institute of High Energy Physics (IHEP), Beijing 100049, China}

\collaboration{The KIMS Collaboration}

\emailAdd{hkpark@ibs.re.kr}

\abstract{The results of a search for solar axions from the Korea Invisible Mass Search (KIMS) experiment 
at the Yangyang Underground Laboratory  are presented.
Low-energy electron-recoil events would be produced by conversion of solar axions  into electrons via the axio-electric effect in  CsI(Tl) crystals.
Using data from an exposure of 34,596 $\rm kg \cdot days$, we set a 90 \% confidence level upper limit on the axion-electron coupling, $g_{ae}$, 
of $1.39 \times 10^{-11}$ for an axion mass less than 1 keV/$\rm c^2$. 
This limit is lower than the indirect solar neutrino bound, and 
fully excludes QCD axions  heavier than 0.48 eV/$\rm c^2$  and 140.9 eV/$\rm c^2$ for the DFSZ and KSVZ models respectively.
 }


\begin{document} 
\maketitle
\flushbottom

\section{Introduction}
\label{sec:intro}
Despite its success, the Standard Model of particle physics still has many problems.
One such problem, known as the  strong CP problem~\cite{hooft}, is that the CP-violating term in strong interaction implies
that the neutron electric dipole moment has to be 
an order of $10^{10}$ larger than the experimental upper bound~\cite{edm}.    
Peccei and Quinn~\cite{pq} found out an elegant method to solve this problem by introducing a new global chiral symmetry $\mathrm{U(1)_{PQ}}$
which is spontaneously broken at an energy scale $f_a$  and which compensates the CP-violating term. 
This solution implies the existence of a new pseudoscalar particle called the axion ($a$)~\cite{weinberg}.  
Since the original axion model  assumed $f_a$ to be at the electroweak energy scale, 
it was ruled out by laboratory experiments~\cite{exp_axion}.  
Currently the invisible axion models with the energy scale $f_a$  as a free parameter, allowing up to the Plank mass scale of $10^{19}$ GeV,  
are not excluded by terrestrial experiments and astrophysics~\cite{pdg}.
There are two popular models, the KSVZ (hadronic)~\cite{KSVZ} and  DFSZ (non-hadronic)~\cite{DFSZ} models.    

\begin{figure}[htp]
\centering 
\includegraphics[width=.6\textwidth]{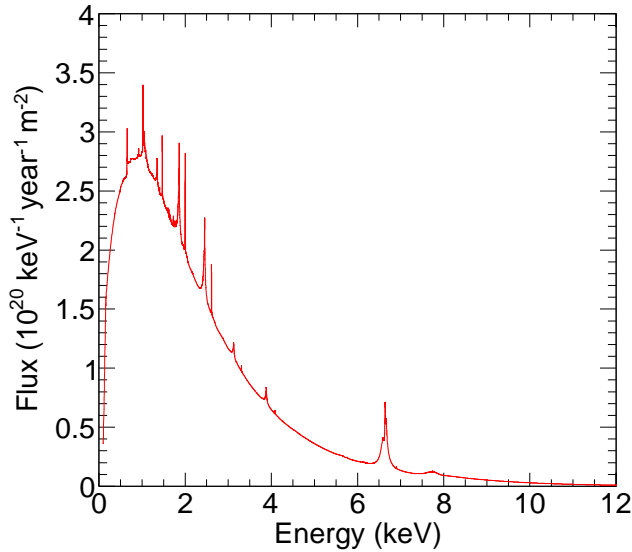}
% "\includegraphics" is very powerful; the graphicx package is already loaded
\caption{\label{fig:flux} Flux of solar axions due to Compton scattering, axio-recombination, axio-deexcitation, 
axio-bremsstrahlung and electron-electron collisions  on earth \cite{axionflux}  with axion-electron coupling of $g_{ae}=10^{-13}$.}
\end{figure}

The strengths of axion-photon ($g_{a\gamma}$), axion-electron ($g_{ae}$) and axion-nucleon ($g_{aN}$) couplings are different for both models  
as described in ref.~\cite{coupling}.  In particular, axion-electron coupling in the DFSZ model occurs at tree level while axion-electron coupling in the KSVZ model 
is  strongly suppressed due to axion-electron coupling at loop level.   
Thus,  in the DFSZ model, the processes related to axion-electron coupling~\cite{comp, recom, axiobrem, eebrem} would prevail
over the Primakoff process with axion-photon coupling  
as an axion production mechanism in stars and the sun: Compton scattering ($\gamma+e \ra e+ a$), 
axio-recombination ($ e + A \ra A^{-} + a $),
axio-deexcitation ($A^{*} \ra A + a $), axio-bremsstrahlung  ($e + A \ra e + A + a$),  and electron-electron collision ($e + e \ra e + e + a$),
where $A$ is an atom. 
The total axion flux on earth produced from the sun was recently estimated in ref.~\cite{axionflux}, which  includes processes
with axion-electron and axion-photon couplings, as shown in figure~\ref{fig:flux}. 

In this paper, we report on a solar axion search using the data sample from the KIMS experiment with CsI(Tl) crystal detectors. 
% The solar axion flux on earth  is recently estimated in ref.~\cite{axionflux}  including processes
% with axion-electron and axion-photon couplings as shown in figure~\ref{fig:flux}. 
Since this estimation in ref.~\cite{axionflux} does not have corrections for axions heavier than
1 keV/$\rm c^2$, our search region for axions  is below this value.

\begin{figure}[htp]
\centering 
\includegraphics[width=.6\textwidth]{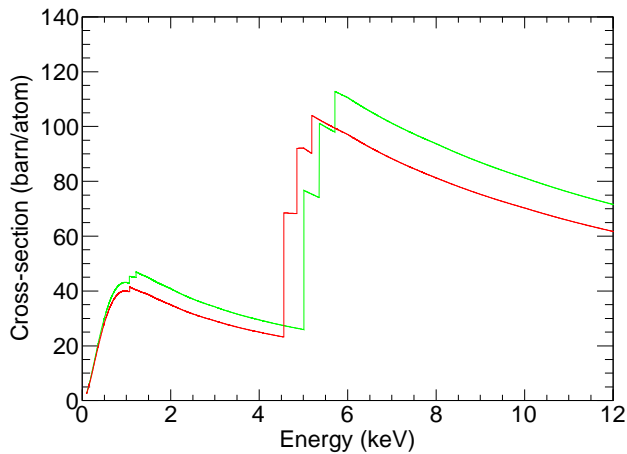}
\caption{\label{fig:crossxectionCsI} Axio-electric cross section calculated for Cs (green ) and I (red) atom for axion mass of 0 keV/$\rm c^2$ with $g_{ae}=1$. }
\end{figure}

Axions would produce electron signals in  the CsI(Tl) detector through the axio-electric effect, $a + A \ra e^- + A^+$ 
where $A$ is mainly either Cs or I in the detector. We searched for this process as a signal for solar axion detection.
The cross section for the axio-electric effect~\cite{axioele} is given by
\begin{equation}
\label{eq:axioele}
\sigma_{ae}(E_a)= \sigma_{pe}(E_a) \frac{g^2_{ae}}{\beta_a} \frac{3E_a^2}{16 \pi \alpha m_e^2}(1- \frac{\beta_a^{\frac{2}{3}}}{3}),
\end{equation}
where $E_a$ is the axion energy, $\sigma_{pe}$ is the photoelectric cross section for either Cs or I in ref.~\cite{photo}, 
$g_{ae}$ is the axion-electron coupling, $\beta_a$ is the axion velocity over the speed of light, 
$\alpha$ is the fine structure constant, and $m_e$ is the electron mass. 
Figure~\ref{fig:crossxectionCsI} shows the cross sections for the axio-electric effect for Cs and I atoms with $g_{ae}=1$.

\section{KIMS Experiment}
The KIMS experiment is  designed to directly search for weakly interacting massive particles (WIMP) 
using CsI(Tl) crystal detectors.  
 The experiment is housed in the Yangyang Underground Laboratory (Y2L) with an earth overburden of 700 m (2400 m water equivalent) 
 and uses a 12 module array of low-background CsI(Tl) crystals with a total mass of 103.4 kg. 
Each detector module is composed of a CsI(Tl) crystal with dimension of 8~cm~x~8~cm~x~30~cm 
and with photomultiplier tubes (PMT)   mounted at each end.
The amplified signals from the PMTs  on each crystal were recorded by a 400 MHz flash analog-to-digital  
converter for a duration of 32 $\mu$s
with the trigger condition requiring at least two photoelectrons (PEs) in both PMTs on each crystal within a 2 $\mu$s window.
The number of PEs are 5 to 6 per keV.
The crystal array is completely surrounded from inside to outside by 10 cm of copper, 5 cm of polyethylene, 15 cm of lead,
and a buffer consisting of liquid scintillator (LS) of 30-cm thickness.
The LS buffer reduces external neutrons and gammas and is equipped with PMT's in order to reject cosmic-ray muon events.
% equipped with PMTs to reduce external neutrons and gammas and to reject cosmic-ray muon events.
The experiment took stable data with 12 crystal modules in the period from September 2009 to December 2012.
Details of the experiment can be found elsewhere~\cite{hslee1,hslee2,cuts}.

\section{Data Analysis}
%YS: describe event selection cuts
This analysis is based on one year data  corresponding to an exposure of   34,596 $\rm kg \cdot days$. 
%which was collected using 11 crystal modules.
% and one crystal with high background was excluded.
We applied event selection criteria that were developed for low-mass WIMPs search studies~\cite{cuts}.
One of the main sources of background events is PMT noise. 
In order to reject these events, a set of event-selection criteria was 
developed by studying noise signals from a dummy detector module consisting of  PMTs mounted on both ends of a transparent and empty acrylic box.
% studied by using a transparent acrylic box (dummy detector) with PMTs mounted at both ends to mimic only PMT noise effect from the real CsI(Tl) detector.    
The dummy detector was operated simultaneously with the CsI(Tl) detector array. 
These event-selection criteria were applied for the recorded events. 
In addition to these criteria, events induced by high-energy cosmic-ray muons  were rejected by 
coincidence with the muon veto detector.

\begin{figure}[htp]
\centering 
\includegraphics[width=.6\textwidth]{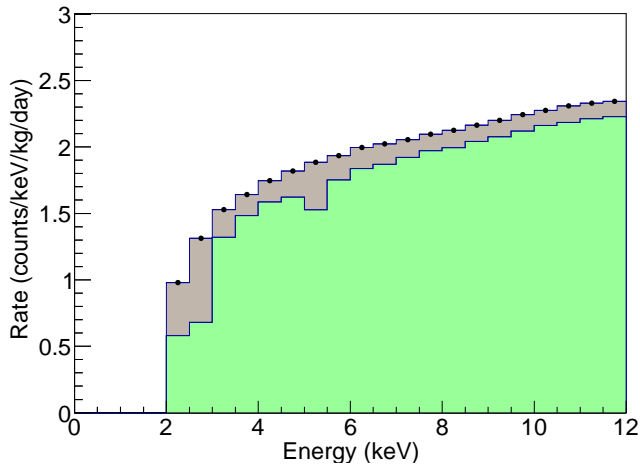}
\caption{\label{fig:fit} Contribution of electron recoil events (green)  and surface alpha events (grey) to the observed single-detector energy spectrum (dots).
 % Composition of each type of event estimated by the fit in the $S$ sample. 
 % The dots are data,  the green area is  the ER event, and the grey area is the SA event. 
}
\end{figure}

Events that passed the above selection criteria were divided into two independent event sets, single-detector ($\rm SD$) and multiple-detector ($\rm MD$) events.   
The $\rm MD$  events are defined as those for which multiple detectors each independently satisfied the trigger condition.
% which have only one  detector-module and at least two detector-modules satisfying trigger condition, respectively.
Since an axion would give rise to an electron-like signal with a hit  in only a single detector-module, 
only $\rm SD$ events were selected as axion candidate events.
% the single-scattering event sample  was used to select axion candidate events. 
The $\rm SD$ events include
surface $\alpha$ events ($\rm S_{\alpha}$) and  
electron recoil events ($\rm R_{e^-}$) from Compton scattered $\gamma$~rays and $\beta$~decays in the crystal bulk~\cite{single}.
% Compton scattered $\gamma$ rays and $\beta$ decays  (ER) events~\cite{single}. 
The $\rm S_{\alpha}$ events come from decays of radioactive isotopes which contaminate the surfaces of the crystals or the surrounding materials.
% particles coming from radioactive isotopes adhered to the crystal surface.  
Major internal backgrounds for $\beta$-decays in our CsI(Tl) crystals are  
$\rm ^{137}Cs$ (Q=1175.6 keV), $\rm ^{134}Cs$ (Q=2058.7 keV) and $\rm ^{87}Rb$ (Q=282 keV). 
The energy spectra from those radioisotopes are flat in our search region, 2 keV to 12 keV,  as 
from Compton-scattered $\gamma$ rays in the MD events~\cite{single}.
% The energy spectra for  $\rm R_{e^-}$ events in the detector is expected to be a flat distribution in the axion search window. 
% The MD events mainly originate from Compton-scattered $\gamma$ rays.
% In the multiple-scattering events ($M$) sample, they are originated from mainly Compton scattered $\gamma$ rays.
Therefore we expect that the MD energy spectrum  is similar to  the $\rm R_{e^-}$ spectrum  in the SD sample.
That is, the energy spectra for  $\rm R_{e^-}$ events in the detector is expected to be a flat distribution in the axion search window.
Pulse-shapes of photoelectron
distributions in the time domain depend on the type of particle incident on the crystal.  
To discriminate $\rm R_{e^-}$ events from $\rm S_{\alpha}$ events we 
employed the pulse-shape discrimination (PSD) method described in refs.~\cite{cuts,psd,mt}.
In this method, the mean time ($MT$) for each event is calculated as follows:
\begin{equation}
 MT = \int t f(t) dt /\int  f(t) dt, \nonumber
\end{equation}
where $f(t)$ is the PE distribution.  
 The quantity (LMT10) is obtained by taking base 10 logarithm of $MT$. 
 % logarithm to base 10 of $MT$ for each type of event 
The LMT10 distribution of each event type is well fitted by an asymmetric gaussian function defined as follows,
\begin{eqnarray}
 g(t)  &=& \frac{A}{1/2(\sigma_L + \sigma_R)} e^{-\frac{1}{2}(\frac{t-\mu}{\sigma_L})^2},~ t < \mu, \nonumber \\  
          &  &  \frac{A}{1/2(\sigma_L + \sigma_R)} e^{-\frac{1}{2}(\frac{t-\mu}{\sigma_R})^2},~ t \ge \mu, \nonumber
\label{eq:asym1}            
\end{eqnarray} 
where $A$ is the amplitude, $\mu$ is the mean value and $\sigma_L$ ($\sigma_R$) is the standard deviation of left (right) side.  
The parameters, $\mu$, $\sigma_L$ and $\sigma_R$, for the $\rm R_{e^-}$  events were 
first determined  from the single-asymmetric-gaussian function fit to the MD sample data. 
In order to extract these fit  parameters for  $\rm S_{\alpha}$ events, we applied fit  to
 the  data  from a sample of a CsI crystal contaminated by $^{222} \rm Rn$ progenies. 
%We performed a fit using the sum of two asymmetric gaussian functions to estimate fractions of SA and ER events 
%for the LMT10 distribution in the $SD$ sample in each 0.5 keV bin.
%From the fit,  the amplitudes for  SA and ER events were estimated with the fixed  parameters,  $\mu$, and $\sigma_L$, $\sigma_R$, 
%for the ER and SA events.
With these parameters fixed, the contributions of  $\rm R_{e^-}$ and $\rm S_{\alpha}$ events in the SD data were determined by the fit and are shown in figure~\ref{fig:fit}. 
%The resulting contribution determined for
%each type of event is shown in figure~\ref{fig:fit}.

In our detector, the expected number of axion events is given by 
\begin{eqnarray}
\label{eq:rate}
 R(E) & = & \int dE_a \frac{d\Phi_a}{dE_a} \epsilon(E)  (\sigma^{Cs}_{ae}(E_a)  N_{Cs} + \sigma^{I}_{ae}(E_a) N_{I}) T R_{det}(E,E_a) \nonumber \\ 
          &\propto  & g_{ae}^4,  
\end{eqnarray}
where $\frac{d\Phi_a}{dE_a}$ is the differential axion flux on the earth, $\epsilon(E)$ is the detection efficiency,
$\sigma^{Cs}_{ae}(E_a)$  and $\sigma^{I}_{ae}(E_a)$ are the axio-electric cross section for Cs and I atoms, respectively, 
$N_{Cs}$ and $N_{I}$ are the number of Cs and I atoms, respectively, in our detector, $T$ is the detector live time, and 
$R_{det}(E,E_a)$ is the resolution function of our detector.

% YS:  describe efficiency
The efficiency,  $\epsilon(E)$, is 
estimated from the ratio of the number of MD events satisfying event selection cuts to the total number of MD events in each energy bin.
The event selection efficiency is energy dependent and varies from 31.0\%  to 91.2\%.

% YS: describe genat4 simulation and figure for data MC comparison for Fe calibration
The resolution function, $R_{det}(E,E_a)$, is determined from a detector simulation.
For each crystal, the photoelectron yield used in the simulation was estimated using  data 
from the 59.4 keV $\gamma$ generated from an $\rm ^{241}Am$ calibration source \cite{cuts,psd}.
% Using a 59.54 keV  $\gamma$ calibration data from $^{241}$Am source, 
% each detector response in the simulation has been tuned by estimating a photoelectron yield of each crystal \cite{cuts,psd}.
%For a given energy, detector response was estimated in number of photoelectron, 
%which was converted to deposited energy in each crystal.

\begin{figure}[htpb]
\centering 
\includegraphics[width=.6\textwidth]{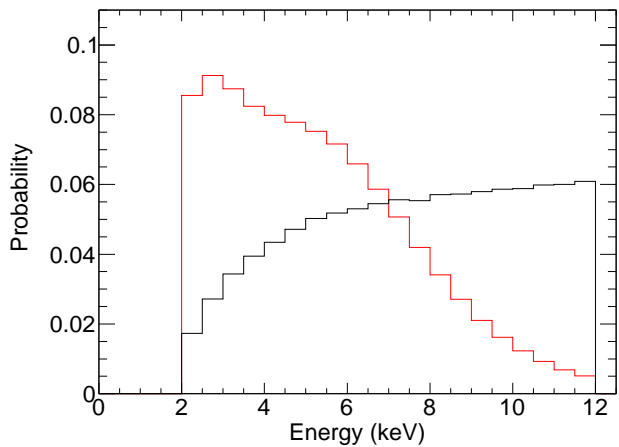}
\caption{\label{fig:pdf} The PDF's for the axion (red) and background events (black).  }
\end{figure}

To estimate the number of axion events, we used the energy spectrum for the  $\rm R_{e^-}$ events in the SD sample,
which contains background events mainly from Compton scattered gamma rays and from $\beta$ decays. 
% consists of axion signal and background, mainly Compton scattered $\gamma$ rays and $\beta$ events.
The signal yield for axion event is extracted by maximizing a binned maximum likelihood function 
for the energy spectrum, which is given by
\begin{equation}
\label{eq:lik}
 \mathcal{L}= \prod_{i=1}^{N_{bin}} e^{-(n_s P_s(E_i) +n_b P_b(E_i))} \frac{ (n_s P_s(E_i) + n_b P_b(E_i))^{N_i}}{N_i !} , \nonumber
\end{equation}
where $N_{bin}$ is the number of bins, $n_s$  and $n_b$ are the expected number of signal and background events, respectively, 
$N_i$  is the number of data events, and $P_s(E_i)$  and $P_b(E_i)$ are 
 the probability density function (PDF) for signal and background in the energy bin $E_i$, respectively. 
The PDF for the energy spectra for the axion signal,  $P_s(E)$, is constructed from the simulation 
by generating electron events with an energy distribution of $R(E)$.
% with the event rate for axion event, $R(E)$, in each energy bin. 
In order to model the background PDF below 12 keV, $P_b(E)$, we used the energy spectrum in the MD sample. 
This is possible because the spectrum contains only a flat Compton continuum, modified by the low-energy efficiency curve.
Figure~\ref{fig:pdf} shows the distributions for $P_s (E)$ and $P_b(E)$.

\begin{figure}[htpb]
\centering 
\includegraphics[width=.6\textwidth]{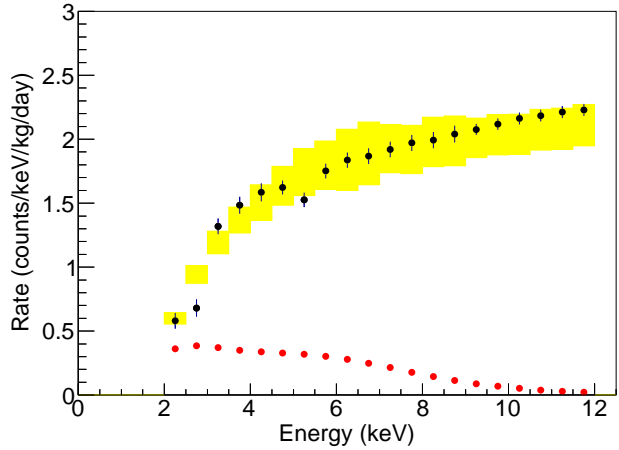}
\caption{\label{fig:fit_back_axion} The energy distributions for the $\rm R_{e^-}$ events in the SD sample (black circle). 
The yellow shaded boxes are the background events estimated by the fit with the efficiency uncertainty.
The red circles are the axion signals scaled up by a factor of million for better visibility.  }
\end{figure}

The signal  yields, $n_s$, for axion masses of 0 keV/$\rm c^2$ to 1 keV/$\rm c^2$ are found to be 
$0.077^{+36.59}_{-127.64}$ to $0.077^{+40.22}_{-132.12}$ 
%$0.072^{+72.4}_{-116.9}$ ($575667^{+1991.34}_{-1664.93}$ )
% to $0.072^{+77.5}_{-323.0}$ ($575672^{+2121.28}_{-2244.6}$)
events/year, consistent with  no axion event. 
Figure~\ref{fig:fit_back_axion} shows the energy distributions for  $\rm R_{e^-}$ events in the SD sample, 
the background events ($\rm R_{e^-}$ events) in the MD sample estimated by the fit and axion signal events.

% \section{Results}
%The extracted signal yield has to be gaussian in a good approximation. 
A 90 \% confidence limit (C.L.) for the signal yield, $n_s^{up}$,  is obtained from
\begin{equation}
\label{eq:cl}
 \frac{\int_0^{n_s^{up}} \mathcal{L} (n_s) d n_s} {\int_0^{\infty} \mathcal{L}(n_s) d n_s}=0.9.
\end{equation}

The resulting values obtained for  $n_s^{up}$ are varied from 58.56 to 60.92 events with axion masses of 0 keV/$\rm c^2$ to 1 keV/$\rm c^2$.
The upper limit on $g_{ae}$ at the 90\% C.L. is estimated with eq.~\ref{eq:rate}, and 
 is found to be $g_{ae} < 1.37  \times 10^{-11}$  and $g_{ae} <  1.39  \times 10^{-11}$ for axion mass of 0 keV/$\rm c^2$ and 1 keV/$\rm c^2$, respectively.
From the upper limit on $g_{ae}$, we exclude a QCD axion heavier than 0.48 eV/$\rm c^2$ in DFSZ model and 140.9 eV/$\rm c^2$ in the KSVZ model. 

\begin{figure}[htpb]
\centering 
\includegraphics[width=.6\textwidth]{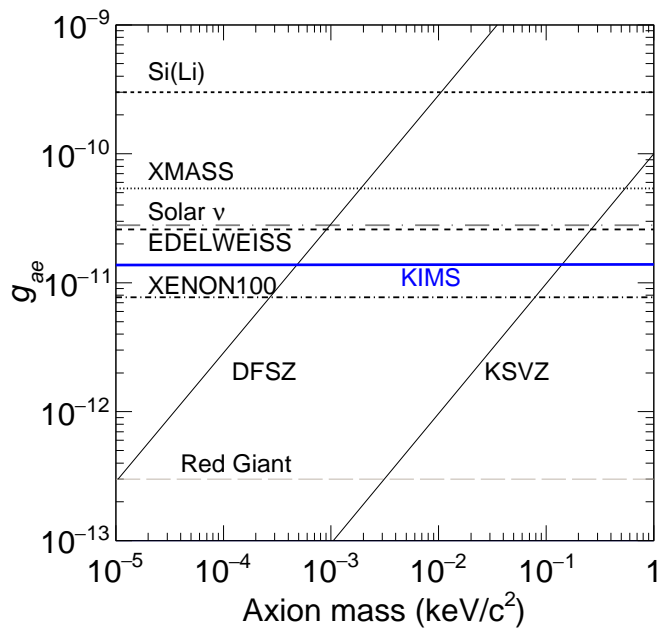}
\caption{\label{fig:gae} The blue line shows the 90 \% C.L. on the axion-electron coupling ($g_{ae}$) for the KIMS.  
The dotted lines are limits by XMASS~\cite{xmass}, EDELWEISS-II~\cite{edelweiss}, XENON100~\cite{xenon} and Si(Li)~\cite{si} experiments. 
The dash-dotted line shows indirect astrophysical bounds, solar neutrino~\cite{solarnu} and red giants~\cite{redgiant}.
The gray lines are predictions by the KSVZ~\cite{KSVZ} and DFSZ~\cite{DFSZ} models.   }
\end{figure}  

\section{Summary}
A search for solar axions from 34,956 $\rm kg \cdot days$ exposure with the KIMS CsI(Tl) detector array has been performed.
In this search, we used the solar axion flux recently estimated with the DFSZ model assuming
 that axions produce electron signals
in the CsI(Tl) detector through the axio-electric effect. The number of extracted axion events is consistent with no axion signal in this data sample.
At the 90 \% C.L.,  we obtain an upper limit of the axion-electron coupling, 
$g_{ae} < 1.39 \times 10^{-11}$ for axion mass of 0 keV to 1 keV 
and  exclude QCD axions heavier than 0.48 eV/$\rm c^2$ in the DFSZ model 
and 140.9 eV/$\rm c^2$ in the KSVZ model.
We exclude a region in the plane of axion mass  and the axion-electron coupling  
at 90 \% C.L. as shown in figure~\ref{fig:gae}.

\acknowledgments
We thank the Korea Midland Power Co. and Korea Hydro and Nuclear Power Co. for providing the underground laboratory space at Yangyang. 
We acknowledge support from the Institute for Basic Science~(IBS) in Korea under the project code IBS-R016-D1, IBS-R016-D2,
the WCU program (R32-10155), 
and the National Research Foundation of Korea (NRF-2011-0031280 and NRF-2011-35B-C00007).

\end{document}